\documentstyle[12pt,epsf]{article}
\newcommand{\beq}{\begin{equation}}
\newcommand{\eeq}{\end{equation}}
\newcommand{\beqa}{\begin{eqnarray}}
\newcommand{\eeqa}{\end{eqnarray}}

\begin{document}

\thispagestyle{empty}
\begin{flushright}
SLAC-PUB-7165\\
FERMILAB-PUB-96/095-T\\
hep-ph/9605259\\
May 1996
\end{flushright}
\vspace*{2cm}
\centerline{\Large\bf The Width Difference in the
            $B_s-\bar B_s$ System
\footnote{
Research supported by the Department of Energy under contract 
DE-AC03-76SF00515.}}
\vspace*{1.5cm}
\centerline{{\sc M. Beneke$^1$}, {\sc G. Buchalla$^2$} 
and {\sc I. Dunietz$^2$}}
\bigskip
\centerline{\sl $^1$Stanford Linear Accelerator Center,}
\centerline{\sl Stanford University, Stanford, CA 94309, U.S.A.}
\vskip0.6truecm
\centerline{\sl $^2$Theoretical Physics Department}
\centerline{\sl Fermi National Accelerator Laboratory}
\centerline{\sl P.O. Box 500, Batavia, IL 60510, U.S.A.}

\vspace*{1.5cm}
\centerline{\bf Abstract}
\vspace*{0.2cm}
\noindent 
We use the heavy quark expansion to investigate
the width difference $\Delta\Gamma_{B_s}$ between the $B_s$ mass 
eigenstates. The corrections of ${\cal O}(\Lambda_{QCD}/m_b)$ and
${\cal O}(m_s/m_b)$ to the leading order expression in the
operator product expansion are derived and estimated to yield
a sizable reduction of the leading result for $\Delta\Gamma_{B_s}$
by typically $30\%$. 
For completeness we also quantify small effects due to
penguin operators and CKM suppressed contributions.
Based on our results we discuss the prediction for
$(\Delta\Gamma/\Gamma)_{B_s}$ with particular emphasis on
theoretical uncertainties. We find
$(\Delta\Gamma/\Gamma)_{B_s}=0.16^{+0.11}_{-0.09}$, where the
large error is dominated by the uncertainty in
hadronic matrix elements. An accuracy of about $10\%$ in
$(\Delta\Gamma/\Gamma)_{B_s}$ should be within reach, assuming
continuing progress in lattice calculations.
In addition we address phenomenological issues and 
implications of a $\Delta\Gamma_{B_s}$ measurement for
constraints on $\Delta M_{B_s}$ and CKM parameters.
We further consider in some detail the lifetime ratio
$\tau(B_s)/\tau(B_d)$ and estimate that, 
most likely, $|\tau(B_s)/\tau(B_d)-1|<1\%$.
\vspace*{0.5cm}
\noindent 

\vspace*{1.2cm}
\noindent
PACS numbers: 13.25.Hw, 14.40.Nd

\vfill

\newpage
\pagenumbering{arabic}

\section{Introduction}
\label{intro}

Mixing phenomena in neutral $B$ meson systems provide an
important testing ground for standard model flavordynamics.
The mass difference between the $B_d$ eigenstates,
$\Delta M_{B_d}$, gave the first evidence for
a large top quark mass and provides a valuable constraint
on $|V_{td}|$ and the CKM unitarity triangle. A direct measurement of
$\Delta M_{B_s}$, the corresponding quantity for $B_s$ mesons, through 
$B_s$-$\bar{B}_s$ oscillations, would  yield further
information and help to reduce hadronic uncertainties
in the extraction of CKM parameters. Complementary
insight can be gained from the width difference
$\Delta\Gamma_{B_s}$ between the $B_s$ mass eigenstates \cite{DUN1,BP}. 
This width difference is expected
to be the largest among bottom hadrons \cite{BIG}, and 
it may be large enough to be accessible by
experiment in the near future. The width difference for $B_d$ mesons, 
on the other hand, is CKM suppressed and experimentally much harder
to determine.

If $\Delta\Gamma_{B_s}$ is indeed found to be sizable, the observation 
of CP violation and the extraction of CKM phases from untagged 
$B_s$ data samples can be contemplated \cite{DUN1,untag,DDF}.
This possibility could be important in two respects. First, tagging 
any $B_s$ data sample costs in statistics and in purity. Second, the 
rapid oscillations dependent on $\Delta M_{B_s} t$ all cancel in
time evolutions of untagged $B_s$ data samples, which are governed
by the two exponentials $\exp(-\Gamma_L t)$ and
$\exp(-\Gamma_H t)$ alone.

The present article continues previous work 
of one of us \cite{DUN1} on the phenomenological potential of 
$\Delta\Gamma_{B_s}$, 
and focuses on theoretical uncertainties and improvements of
the prediction.
We compute the width difference in the heavy quark 
expansion and include explicit $1/m_b$-corrections, which 
improves over previous estimates of $\Delta \Gamma_{B_s}$ based on a 
partonic \cite{HAG,FLP,CHA,BSS,VUKS} or exclusive \cite{ALE93} 
approach and allows us to assess the remaining uncertainties more 
reliably. Combined with future measurements of $\Delta \Gamma_{B_s}$ 
these predictions can be used to derive indirect constraints 
on $|V_{ts}/V_{td}|$ \cite{BP} and $\Delta M_{B_s}$. 
Non-standard model sources of CP violation in the $B_s$ system 
would reduce $\Delta \Gamma_{B_s}$ compared 
to its standard model value, as explained in \cite{GRO96}, so 
that a lower bound on the standard model prediction is  
especially interesting.

Starting from the flavor eigenstates 
$\{|B_s\rangle,\, |\bar B_s\rangle\}$, $B_s-\bar{B_s}$ mixing is 
determined by the $2\times 2$ matrix 

\begin{equation}
{\bf \cal  M} = {\bf M}-\frac{i}{2}\,{\bf \Gamma}.
\end{equation}

\noindent with hermitian ${\bf M}$ and ${\bf\Gamma}$.
Due to CPT conservation $M_{11}=M_{22}\equiv M_{B_s}$, 
$\Gamma_{11}=\Gamma_{22}\equiv\Gamma_{B_s}$.
We recall that for the $B_s$ system the off-diagonal elements
obey the pattern

\begin{equation}
\left|\frac{\Gamma_{12}}{M_{12}}\right|\sim {\cal O}\!\left(\frac{m_b^2}
{m_t^2}\right).
\end{equation}

\noindent The mass and lifetime 
difference between eigenstates are given by (`H' for `heavy', 
`L' for `light')

\begin{equation}\label{delmex}
\Delta M_{B_s} \equiv M_H-M_L=2 \,|M_{12}|,\\[0.1cm]
\end{equation}
\begin{equation}\label{delgex}
\Delta \Gamma_{B_s} \equiv \Gamma_L-\Gamma_H = -\frac{2\,\mbox{Re}\,(
M_{12}^*\Gamma_{12})}{|M_{12}|}.
\end{equation} 

\noindent The corrections to (\ref{delmex}) and (\ref{delgex})
are extremely suppressed. They enter only at order
$|\Gamma_{12}/M_{12}|^2$ and vanish altogether in the limit of
exact CP symmetry.
Anticipating the actual hierarchy of eigenvalues, 
we have defined both $\Delta M_{B_s}$ and $\Delta\Gamma_{B_s}$ 
to be positive quantities.

Neglecting CP violating corrections, which are very small in the
standard model (SM), the mass eigenstates are CP eigenstates (up to 
corrections of at most $10^{-3}$), 
and with the phase convention 
$CP|B_s\rangle=-|\bar B_s\rangle$ one has 
$|B_{H/L}\rangle=(|B_s\rangle\pm |\bar B_s\rangle)/\sqrt{2}$. 
Then\footnote{
Subsequently, we present the result of our calculation of $\Gamma_{21}$ 
as a result for $\Delta\Gamma_{B_s}$ using (\ref{delgacp}). If one does 
not want to assume standard model CP violation, 
(\ref{delgacp}) must be generalized to (\ref{delgex}),
but our result for $\Gamma_{21}$ is still 
valid, provided non-standard model CP violation modifies only 
$M_{12}$, but not $\Gamma_{12}$. Since $\Gamma_{12}$ results 
predominantly from tree 
decays, this is reasonable to assume.},
using standard CKM phase conventions \cite{PDG},  

\begin{equation}
\label{delgacp}
\Delta \Gamma_{B_s}=-2\,\Gamma_{12}=-2\,\Gamma_{21}.
\end{equation}

\noindent Note that the lighter state is CP even \cite{DUN1} 
and decays more rapidly 
than the heavier state. This also follows from the fact that most of the 
decay products in the $b\to c\bar{c}s$ transition which are common to 
$B_s$ and $\bar{B_s}$ are CP even \cite{ALE93}.

Both the mass and lifetime difference are determined by the familiar 
box diagrams that give rise to an effective $\Delta B=2$ 
Hamiltonian (`$B$' denotes $b$-quark number). On distance scales 
larger than $1/M_W$, but still smaller than $1/m_b$, this effective 
Hamiltonian contains a local $\Delta B=2$ interaction as well as 
a bilocal part constructed from two (local) $\Delta B=1$ 
transitions. The mass difference is given by the real part  
of the box diagram and is dominated by the top quark contribution. 
For this reason, $M_{12}$ 
is generated by an interaction that is local already on scales 
$x>1/M_W$ and theoretically well under control.
The short-distance contribution has been calculated to next-to-leading
order in QCD \cite{BJW}. The long-distance contribution is parametrized 
by the matrix element of a single four-quark operator between
$B_s$ and $\bar B_s$ states. Corrections to this result are 
suppressed by powers of $m^2_b/M^2_W$ 
and completely irrelevant for all practical purposes.

The lifetime difference is given by the imaginary part of the 
box diagram and determined by real intermediate states,
which correspond to common decay products 
of $B_s$ and $\bar{B}_s$, so that only the 
bilocal part of the $\Delta B=2$ Hamiltonian can contribute. 
The presence of long-lived (on hadronic scales) intermediate states 
would normally preclude a short-distance treatment of the lifetime 
difference as indeed it does for neutral kaons. But for bottom mesons, 
the $b$ quark mass $m_b$ provides an additional short-distance 
scale that leads to a large energy release 
(compared to $\Lambda_{QCD}$) into the intermediate states. Thus, 
at typical hadronic distances $x > 1/m_b$, the decay is again a 
local process. The bilocal $\Delta B=2$ Hamiltonian can be expanded 
in inverse powers of the heavy quark mass, schematically:

\begin{equation}
\label{bilocalexpand}
\mbox{Im}\,i\int d^4 x\,T\!\left({\cal O}^{\Delta B=1}(x) {\cal O}^{
\Delta B=1}(0)\right) = \sum_n\frac{C_n}{m_b^n}\,{\cal O}_n^{\Delta B=2}(0)
\end{equation}

\noindent The matrix elements of local $\Delta B=2$ operators 
that appear here and in the mass difference are not independent 
of $m_b$. Their mass dependence could be made explicit with the 
help of Heavy Quark Effective Theory. The difference between the 
mass and lifetime difference is that for the lifetime difference 
explicit $1/m_b$ corrections arise from the 
expansion (\ref{bilocalexpand}) even before expanding the matrix 
elements of local operators. The heavy quark expansion applies as 
well to the diagonal elements 
$\Gamma_{ii}\equiv\Gamma_{B_s}\equiv(\Gamma_H+\Gamma_L)/2$ 
and has been used 
to predict the total width of bottom hadrons \cite{BIG}. A 
contribution to $\Gamma_{12}$ requires that the spectator strange 
quark and the bottom quark come together within a distance $1/m_b$ 
in a meson of size $1/\Lambda_{QCD}$. This volume suppression 
together with the phase space enhancement, leads to the 
estimate

\begin{equation}\label{estimate}
\left|\frac{\Gamma_{12}}{\Gamma_{11}}\right|\sim 16\pi^2\left(
\frac{\Lambda_{QCD}}{m_b}\right)^3.
\end{equation}

\noindent The application of heavy quark expansions to non-leptonic 
decays assumes local duality. The accuracy of this assumption can 
not be quantified within the framework itself, at least not to finite order 
in the heavy quark expansion. The assumption that the sum over 
exclusive modes is accurately described by the heavy quark expansion 
might be especially 
troubling for $\Delta \Gamma_{B_s}$, since it is saturated by only 
a few $D_s^{(*,**)}\bar{D}_s^{(*,**)}$ intermediate states and the energy 
release is only slightly larger than one GeV. On the other hand, in the 
small-velocity limit $\Lambda_{QCD}\ll m_b-2 m_c\ll m_c$, and the
$N_c\to\infty$-limit\footnote{This limit is necessary to justify the 
factorization assumption for four-fermion operators.}, local duality 
with only a few intermediate states can indeed be verified 
explicitly \cite{ALE93}. 

This article starts from the hypothesis that duality violations should 
be less than $10\%$ for $\Delta \Gamma_{B_s}$. 
Aiming at an accuracy of $10\%$, the following corrections 
to the leading order result have to be considered:

\renewcommand{\labelenumi}{(\roman{enumi})}
\begin{enumerate}
\item $1/m_b$ corrections from dimension seven operators in 
(\ref{bilocalexpand}).
\item Deviations from the `vacuum insertion' (`factorization') 
assumption for matrix 
elements of four-fermion operators.
\item Radiative corrections of order $\alpha_s/\pi$.
\item Penguin and Cabibbo-suppressed contributions.
\end{enumerate}

\noindent The major part of this paper is devoted to $1/m_b$ corrections. 
We hope to return to radiative corrections in a subsequent publication.
These would bring the short-distance part of the calculation for
$\Delta\Gamma_{B_s}$ on the same level that has already been
achieved for $\Delta M_{B_s}$.
The result for $\Delta \Gamma_{B_s}$ to next-to-leading order in the 
$1/m_b$ expansion is obtained in Sect.~\ref{basic}. We use the vacuum 
insertion approximation for the dimension seven operators, and express 
the result in terms of two non-perturbative parameters that have to 
be computed with lattice methods. Sect. \ref{pheno} is devoted to the
phenomenology of $\Delta \Gamma_{B_s}$. Numerical results are 
discussed in Sect.~\ref{numeric}, together with the theoretical 
uncertainties in $\Delta\Gamma_{B_s}/\Gamma_{B_s}$.
In Sect. \ref{limit} a generally valid upper bound on 
$\Delta \Gamma_{B_s}$ is derived. Sect. \ref{measure} describes
potential strategies to measure the width difference in experiment.
Some phenomenological applications of such a measurement are
considered in Sect. \ref{bbmix}.

An issue related to $\Delta \Gamma_{B_s}$ concerns 
the total decay rate $\Gamma_{B_s}$ of $B_s$ mesons, averaged
over the long-lived and short-lived component. For experimental
investigations of $\Delta\Gamma_{B_s}$ \cite{DUN1} it would be
helpful to know to what extent the average $B_s$ decay rate
$\Gamma_{B_s}$ differs from $\Gamma_{B_d}$. These decay widths are
estimated to coincide to a high accuracy \cite{BIG}. We 
quantify this expectation and detail the contributions that could
give rise to a difference between $\Gamma_{B_s}$ and
$\Gamma_{B_d}$ in Sect.~\ref{bsbd}.  
A summary is presented in Sect.~\ref{summary}. 
Penguin and Cabibbo-suppressed contributions turn out to shift 
$\Delta \Gamma_{B_s}$ by less then $10\%$ and are discussed in 
the Appendices, along with a comment
on the lifetime ratio of $B^+$ to $B_d$ mesons.
 
\section{$\Delta\Gamma_{B_s}$ -- Basic Formalism}
\label{basic}

The optical theorem relates the total decay width of a particle
to its forward scattering amplitude. The off-diagonal element 
$\Gamma_{21}$ of the decay width matrix is given by

\begin{equation}\label{g21t}
\Gamma_{21}=\frac{1}{2M_{B_s}}
\langle\bar B_s|{\cal T}|B_s\rangle .
\end{equation}

\noindent The normalization of states is 
$\langle B_s|B_s\rangle=2EV$ (conventional relativistic normalization)
and the transition operator ${\cal T}$ is defined by

\begin{equation}\label{tdef}
{\cal T}=\mbox{Im}\,i\int d^4x\ T\,{\cal H}_{eff}(x){\cal H}_{eff}(0).
\end{equation}

\noindent Here ${\cal H}_{eff}$ is the low energy effective weak
Hamiltonian mediating bottom quark decay. The component that is
relevant for $\Gamma_{21}$ reads explicitly

\begin{equation}\label{heff}
{\cal H}_{eff}=\frac{G_F}{\sqrt{2}}V^*_{cb}V_{cs}
\left(C_1(\mu) (\bar b_ic_j)_{V-A}(\bar c_js_i)_{V-A}+
C_2(\mu) (\bar b_ic_i)_{V-A}(\bar c_js_j)_{V-A}\right),
\end{equation}

\noindent where we are neglecting Cabibbo suppressed channels 
and the contributions from penguin operators, whose coefficients are
small numerically. These contributions will be considered in the 
Appendices. We use the notation $(\bar{q}_1 q_2)_{V-A}=\bar q_1 
\gamma_\mu (1-\gamma_5) q_2$ and similar notation for other 
combinations of Dirac matrices. The indices $i,j$ refer to color.  
The Wilson coefficient functions $C_{1,2}$
read in the leading logarithmic approximation 

\begin{equation}\label{c21pm}
C_{2,1}=\frac{C_+\pm C_-}{2}\qquad
C_+(\mu)=\left[\frac{\alpha_s(M_W)}{\alpha_s(\mu)}\right]^{6/23}\quad
C_-(\mu)=\left[\frac{\alpha_s(M_W)}{\alpha_s(\mu)}\right]^{-12/23}
\end{equation}

\noindent with scale $\mu$ of order $m_b$.

\begin{figure}[t]
   \vspace{0cm}
   \epsfysize=4.8cm
   \epsfxsize=6cm
   \centerline{\epsffile{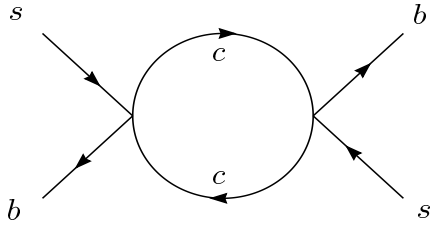}}
   \vspace*{-1.5cm}
\caption{\label{fig1} Diagram that gives the leading and 
next-to-leading in $1/m_b$ terms in the heavy quark expansion of 
the forward scattering amplitude.} 
\end{figure} 

The leading contribution to the $\Delta B=2$ transition operator   
is shown in Fig.~\ref{fig1}, 
where the vertices correspond to the
interaction terms in (\ref{heff}). The operator product 
expansion is constructed using standard methods \cite{BIG}. 
Because of the large momentum flowing through the fermion loop, 
it can be contracted to a point. To leading order in $1/m_b$, 
the strange momentum can be neglected and the $b$ quark momentum 
identified with the meson momentum. The result can be 
expressed in terms of two dimension six operators

\vspace*{-0.2cm}
\begin{eqnarray}\label{qqs}
Q &=& (\bar b_is_i)_{V-A}(\bar b_js_j)_{V-A}\\[0.1cm]
Q_S\!\! &=& (\bar b_is_i)_{S-P}(\bar b_js_j)_{S-P} .
\end{eqnarray}

\noindent The first operator coincides with the single operator that 
contributes to the mass difference. The appearance of a second operator 
can be traced to the fact that in the calculation of $\Gamma_{21}$ 
the external $b$ momentum can not be neglected, because its zero 
component (in the meson rest frame) provides the large momentum scale.

To include $1/m_b$ corrections, the forward scattering amplitude, evaluated 
between on-shell quark states, is 
expanded in the small strange quark momentum and matched onto operators 
with derivatives or with a factor of $m_s$, the strange quark mass, which 
we count as $\Lambda_{QCD}$. Operators with additional gluon fields 
contribute only to corrections of order $(\Lambda_{QCD}/m_b)^2$ and 
need not be considered. It is more direct (and rather trivial at this order) 
to use the background field method \cite{NOV}. Since we do not scale out 
the `kinematic' part of order $m_b$ in derivatives acting on $b$ fields, 
we do not have immediate power counting. Some operators of higher 
dimension in (\ref{bilocalexpand}) have to be kept, if they contain 
derivatives on $b$ fields, such as $R_2$ below. Using the equations 
of motion, we are left with operators with at most one derivative 
on $b$ fields and obtain

\begin{eqnarray}\label{tres}
\Gamma_{21} &=& -\frac{G^2_F m^2_b}{12\pi (2 M_{B_s})}(V^*_{cb}V_{cs})^2
 \sqrt{1-4z}\cdot  \nonumber\\
&&\,\cdot\left[\left((1-z)K_1+\frac{1}{2}(1-4z)K_2\right)\langle Q 
\rangle +
 (1+2z)\left(K_1-K_2\right) \langle Q_S\rangle  + \hat{\delta}_{1/m}\right], 
\end{eqnarray}

\noindent where $z=m^2_c/m^2_b$ and

\begin{equation}\label{k1k2}
K_1=N_c C^2_1+2C_1 C_2\qquad K_2=C^2_2 .
\end{equation}

\noindent The brackets denote the matrix 
element of an operator ${\cal O}$ 
between a $\bar{B}_s$ and $B_s$ state, 
$\langle{\cal O}\rangle\equiv\langle\bar B_s|{\cal O}|B_s\rangle$.
The $1/m_b$ corrections are summarized in 

\begin{eqnarray}\label{oneoverm}
\hat{\delta}_{1/m} &=& (1+2 z)\left[K_1\,(-2\langle R_1\rangle 
-2\langle R_2\rangle) + K_2\,(\langle R_0\rangle -2\langle \tilde{R}_1
\rangle -2\langle \tilde{R}_2\rangle)\right]\nonumber\\
&&\,-\,\frac{12 z^2}{1-4 z}\left[K_1\,(\langle R_2\rangle 
+2\langle R_3\rangle) + K_2\,(\langle \tilde{R}_2\rangle 
+2\langle \tilde{R}_3\rangle)\right] .
\end{eqnarray}

\noindent The subdominant operators are denoted by $R_i$ and 
$\tilde{R_i}$ and read ($R_4$ will be needed below)

\begin{eqnarray}
\label{r0qt}
R_0&=&Q_S+\tilde Q_S+\frac{1}{2}Q\qquad
\tilde Q_S=(\bar b_is_j)_{S-P}(\bar b_js_i)_{S-P}\\
\label{rrt1}
R_1&=&\frac{m_s}{m_b}(\bar b_is_i)_{S-P}(\bar b_js_j)_{S+P}\\
\label{rrt2}
R_2&=&\frac{1}{m^2_b}(\bar b_i {\overleftarrow D}_{\!\rho}
\gamma^\mu(1-\gamma_5)
D^\rho s_i)( \bar b_j\gamma_\mu(1-\gamma_5)s_j)\\
\label{rrt3}
R_3&=&\frac{1}{m^2_b}(\bar b_i{\overleftarrow D}_{\!\rho}
(1-\gamma_5)D^\rho s_i) (\bar b_j(1-\gamma_5)s_j)\\
\label{rrt4}
R_4&=&\frac{1}{m_b}(\bar b_i(1-\gamma_5)iD_\mu s_i)
(\bar b_j\gamma^\mu(1-\gamma_5)s_j).
\end{eqnarray}

\noindent The $\tilde{R}_i$ denote the color-rearranged operators 
that follow from the expressions for $R_i$ by interchanging $s_i$ and 
$s_j$. In deriving (\ref{tres}) we omitted total derivative terms, 
because four-momentum is conserved in the
forward scattering amplitude. 

The operators $R_i$ and $\tilde{R}_i$ are not all independent at order 
$1/m_b$. Relations can be derived by using the equations of motion 
and omitting total derivatives. To reduce $R_0$, one can start from 
the Fierz identity

\begin{eqnarray}\label{fierz}
(\bar{b}_i\gamma_\mu(1-\gamma_5)s_i)(\bar{b}_j\gamma_\nu
(1-\gamma_5)s_j) &=& \\[0.1cm]
&&\hspace*{-4cm} - (\bar{b}_i\gamma_\mu(1-\gamma_5)s_j)
(\bar{b}_j\gamma_\nu(1-\gamma_5)s_i)
+ \frac{1}{2}g_{\mu\nu}\,(\bar{b}_i\gamma^\lambda(1-\gamma_5)s_j)
(\bar{b}_j\gamma_\lambda(1-\gamma_5)s_i)\nonumber
\end{eqnarray}

\noindent and apply derivatives in an appropriate way. Up to 
corrections of $1/m_b$ (or less), we find 

\begin{eqnarray}
\label{relations}
R_0 &=& 2 \tilde{R_1}-R_2+2 R_4\nonumber\\[0.1cm]
\tilde{R}_0 &=& R_0\nonumber\\[0.1cm]
\tilde{R}_2 &=& -R_2\\[0.1cm]
\tilde{R}_3 &=& R_3+R_2/2\nonumber\\[0.1cm]
\tilde{R_4} &=& R_4+\tilde{R_1}-R_1-R_2 .\nonumber
\end{eqnarray}

\noindent The first of these relations shows explicitly that 
the matrix element of $R_0$ is $1/m_b$ suppressed compared to 
$Q$, which is not directly evident from its definition above. 

At this point, we have expressed the $1/m_b$ corrections 
to $\Delta \Gamma_{B_s}$ in terms 
of five new unknown parameters, in addition to the two non-perturbative 
parameters that appear already at leading order, and which also contain 
implicit $1/m_b$ corrections. In principle they can all be obtained 
within the framework of lattice gauge theory\footnote{The matrix elements 
of the subleading operators could be evaluated in the static limit. 
However, to consistently include all $1/m_b$ corrections, 
$\langle Q\rangle$ and $\langle Q_S\rangle$ must be computed either 
in full QCD or in Heavy Quark Effective Theory including $1/m_b$ 
corrections to the Lagrangian as well as to the effective
theory operators. The 
parametrization of $1/m_b$ corrections to $\langle Q\rangle$ has 
been analyzed in \cite{KIL93}.}. 
Unfortunately, results accurate to $10\%$ are not
yet available, especially not for $\langle Q_S\rangle$ (and all the 
subleading operators). We therefore adopt the following strategy: we 
parametrize the two operators that appear at leading order. They 
can be estimated in vacuum insertion or the large $N_c$ limit, but 
should ultimately be computed on the lattice. 
The operators $R_i$, $\tilde R_i$, on the other hand, are only of
subleading importance and we shall content ourselves here with the
factorization approximation.

Following standard conventions we express the matrix elements of
$Q$ and $Q_S$ in terms of the corresponding `bag' parameters $B$
and $B_S$

\begin{eqnarray}\label{meqs}
\langle Q\rangle &=& f^2_{B_s}M^2_{B_s}2\left(1+\frac{1}{N_c}\right)B,
\\[0.1cm]
\langle Q_S\rangle\!\! &=& -f^2_{B_s}M^2_{B_s}
\frac{M^2_{B_s}}{(m_b+m_s)^2}\left(2-\frac{1}{N_c}\right)B_S,
\end{eqnarray}

\noindent where $M_{B_s}$ and $f_{B_s}$ are the mass and decay constant
of the $B_s$ meson and $N_c$ is the number of colors. 
The parameters $B$ and $B_S$ are defined such that 
$B=B_S=1$ corresponds to the
factorization (or `vacuum insertion') approach, which can provide
a first estimate. Factorization of four-fermion operators 
is a controlled approximation 
only for large $N_c$ or for a non-relativistic system. 
In the large $N_c$ limit, $B=3/4$ and $B_S=6/5$. 
In the sense of these limiting cases, factorization 
for realistic $B_s$ mesons 
can be expected to yield the correct order of magnitude and, in 
particular, the right sign of these matrix elements. 
Existing nonperturbative calculations like lattice simulations
for $\langle Q\rangle$, and for its counterpart in the
$K-\bar K$ system, are in agreement with this expectation. 
Beyond these limits factorization does not reproduce the 
correct renormalization scale and scheme dependence, 
necessary to cancel the
corresponding, unphysical dependences in the Wilson coefficients. 
This raises the additional question, to which we return 
below, at what scale factorization should be employed to
estimate the matrix elements. 
Without further information a certain variation
of the parameters $B$, $B_S$ should be allowed in performing
a numerical analysis. 

Next we consider the subleading operators $R_i$, $\tilde{R_i}$,
where we apply factorization.
Using relations such as ($\alpha,\beta$ refers to 
spinor indices, $i,j$ to color as before)

\begin{equation}
\langle\bar{B}_s|\bar b_{\alpha i} {\overleftarrow D}_{\!\rho}
D^\rho s_{\beta j}|0\rangle = \frac{1}{2} (m_b^2-M_{B_s}^2)\, 
\langle\bar{B}_s|\bar b_{\alpha i} s_{\beta j}|0\rangle , 
\end{equation}

\noindent valid to first order in $1/m_b$, all 
matrix elements can be expressed in terms of $f_{B_s}$,
$M_{B_s}$ and quark masses. We find 

\begin{eqnarray}
\langle R_0\rangle &=& f^2_{B_s}M^2_{B_s}\left(1+\frac{1}{N_c}\right)
\left(1-\frac{M^2_{B_s}}{(m_b+m_s)^2}\right)\nonumber\\[0.1cm]
\langle R_1\rangle &=& f^2_{B_s}M^2_{B_s}\frac{m_s}{m_b}
 \left(2+\frac{1}{N_c}\right)\nonumber\\[0.1cm]
\langle\tilde R_1\rangle &=& f^2_{B_s}M^2_{B_s}\frac{m_s}{m_b}
 \left(1+\frac{2}{N_c}\right)\nonumber\\
\label{mer2}
\langle R_2\rangle &=& f^2_{B_s}M^2_{B_s}
 \left(\frac{M^2_{B_s}}{m^2_b}-1\right)
 \left(-1+\frac{1}{N_c}\right)\\[0.1cm]
\langle\tilde R_2\rangle &=& f^2_{B_s}M^2_{B_s}
 \left(\frac{M^2_{B_s}}{m^2_b}-1\right)
 \left(1-\frac{1}{N_c}\right)\nonumber\\[0.1cm]
\langle R_3\rangle &=& f^2_{B_s}M^2_{B_s}
 \left(\frac{M^2_{B_s}}{m^2_b}-1\right)
 \left(1+\frac{1}{2N_c}\right)\nonumber\\[0.1cm]
\langle\tilde R_3\rangle &=& f^2_{B_s}M^2_{B_s}
 \left(\frac{M^2_{B_s}}{m^2_b}-1\right)
 \left(\frac{1}{2}+\frac{1}{N_c}\right).\nonumber
\end{eqnarray}

Combining the above results, one can obtain $\Delta\Gamma_{B_s}$
from (\ref{tres}). The sensitivity to $V_{cb}$ may be eliminated
by normalizing to the total decay rate $\Gamma_{B_s}$ expressed
in terms of the semileptonic width and branching ratio

\begin{equation}\label{gbsl}
\Gamma_{B_s}=\frac{\Gamma(B_s\to Xe\nu)}{B(B_s\to Xe\nu)}=
\frac{G^2_Fm^5_b}{192\pi^3}|V_{cb}|^2
\frac{g(z)\,\tilde{\eta}_{QCD}}{B(B_s\to Xe\nu)},
\end{equation}

\begin{equation}\label{gzdef}
g(z)=1-8z+8z^3-z^4-12z^2\ln z,
\end{equation}

\noindent where 
$B(B_s\to Xe\nu)$ is to be taken from experiment\footnote{Since 
we show in Sect.~\ref{bsbd} that the lifetime difference between 
$B_s$ and $B_d$ is tiny, no attention has to be paid to the 
flavor content of the $B$ meson.} and $z=m^2_c/m^2_b$ as before. 
$\tilde{\eta}_{QCD}$ denotes the one-loop QCD corrections ($m_b$ refers 
to the $b$-quark pole mass). Their analytic 
expression can be found in \cite{NIR}. At $m_b=4.8\,$GeV, 
$m_c=1.4\,$GeV, $\mu=m_b$, and with $\alpha_s(m_b)=0.216$
one has $\tilde{\eta}_{QCD}=0.88$. Since 
radiative corrections to $\Delta \Gamma_{B_s}$ are not yet known, 
the inclusion of radiative corrections to the semileptonic width 
seems somewhat arbitrary. On the other hand, with $V_{cb}=0.04$ and 
$\Gamma^{-1}_{B_s}=1.54\,$ps, one obtains $m_b\approx 4.8\,$GeV from 
(\ref{gbsl}), compared to $m_b\approx 4.5\,$GeV without QCD 
corrections. We prefer the first value as our central choice for 
$m_b$ in the numerical analysis, 
but repeat that, in the absence of radiative corrections 
to $\Delta \Gamma_{B_s}$, $\tilde{\eta}_{QCD}$ can as well be 
considered as a normalization uncertainty that replaces the 
normalization uncertainty due to the errors in $V_{cb}$ and 
$\Gamma_{B_s}$. Finally one arrives at the following expression: 

\begin{eqnarray}\label{dgres}
\frac{\Delta\Gamma_{B_s}}{\Gamma_{B_s}} &=& 16\pi^2
B(B_s\to X e\nu)\frac{\sqrt{1-4z}}{g(z)\,\tilde\eta_{QCD}}
\frac{f^2_{B_s}M_{B_s}}{m^3_b} \,V^2_{cs} \cdot \nonumber \\
&& \cdot \,\Bigg[\left(2(1-z)K_1+(1-4z)K_2\right)
\left(1+\frac{1}{N_c}\right)B \\
&&\hspace*{0.3cm}+\,(1+2z)\left(K_2-K_1\right)
 \frac{M^2_{B_s}}{(m_b+m_s)^2}
 \left(2-\frac{1}{N_c}\right)B_S\,+\, \delta_{1/m} + \delta_{rem}\Bigg]. 
\nonumber
\end{eqnarray}

\noindent 
$\delta_{1/m}$ is related to $\hat{\delta}_{1/m}$, defined in 
(\ref{oneoverm}), through 

\begin{equation}
\hat{\delta}_{1/m} = f_{B_s}^2 M_{B_s}^2 \delta_{1/m} ,
\end{equation}

\noindent and from now on we imply that (\ref{mer2}) is used. 
We have indicated by $\delta_{rem}$ the contributions from CKM-suppressed 
intermediate states $(u\bar{c}, \bar uc, u\bar{u})$ and from penguin 
operators in the $\Delta B=1$ effective Hamiltonian, which are 
estimated in the Appendices A and B 
to be below $\pm 3\%$ and about $-5\%$, 
respectively, relative to the leading order contribution.
We shall neglect $\delta_{rem}$ in the analysis to follow.

Since $f_{B_s}\sim \Lambda_{QCD}^{3/2}/m_b^{1/2}$, $\Delta\Gamma_{B_s}/
\Gamma_{B_s}\sim 16\pi^2 (\Lambda_{QCD}/m_b)^3$ as in the estimate 
(\ref{estimate}). Eq.~(\ref{dgres}) is valid to leading 
(${\cal O}(1/m^3_b)$) and next-to-leading order 
(${\cal O}(1/m^4_b)$) in the heavy quark expansion. The most important 
neglected terms are radiative corrections of order ${\cal O}(\alpha_s/
m_b^3)$. Implicit here is the assumption that the quantity 
$(\Delta\Gamma/\Gamma)_{B_s}$ 
can indeed be represented to reasonable accuracy by the series in powers 
of $\Lambda_{QCD}/m_b$ that is generated by the heavy quark
expansion. As mentioned earlier, this assumption is equivalent to 
the assumption of local quark hadron duality. 

The leading term in (\ref{dgres}), represented by the 
contributions proportional to $B$ and $B_S$, agrees with the results
that have been given previously in the literature\footnote{Often  
factorization is assumed for the leading order term, so that 
$B$ and $B_S$ have 
to be set to unity to recover the result.} \cite{HAG,FLP,CHA,BSS,VUKS}. 
Note that we have consistently kept the distinction between quark masses,
arising from the short-distance loops or the equations of
motion, and the meson mass $M_{B_s}$ from hadronic matrix elements,
since we are aiming at effects beyond leading order in the
heavy quark expansion. 

In (\ref{dgres}), $K_1,K_2$ and $B,B_S$ should be evaluated at a 
scale of order $m_b$. If we wanted to use vacuum insertion to estimate 
the bag factors, it is physically clear, especially in the heavy 
quark limit $m_b\to \infty$, that vacuum insertion should be applied not 
at the scale $m_b$, but at a typical hadronic scale $\mu_h\sim 1\,$GeV. 
This still leaves us with an ambiguity as to the choice of $\mu_h$ 
and in addition with the question, how $B(\mu_h)=B_S(\mu_h)=1$ are 
related to $B(m_b)$ and $B_S(m_b)$. This latter question can be 
answered in the limit $\mu_h\ll m_b$ and corresponds to 
the inclusion of `hybrid logarithms' \cite{VS,PW}, as done in 
\cite{VUKS}. The evolution from $m_b$ to $\mu_h$ is performed in the
leading logarithmic approximation in the static theory and leads 
to\footnote{We have checked the calculation of hybrid logarithms and 
agree with the findings of \cite{VUKS}.}

\begin{eqnarray}\label{bhlog}
B(m_b) &=& 1\\[0.1cm] 
B_S(m_b)\!\!&=& 1-\frac{3}{5}\left(1-\left[
 \frac{\alpha_s(m_b)}{\alpha_s(\mu_h)}\right]^{8/25}\right). \nonumber
\end{eqnarray}

\noindent  The first equation in (\ref{bhlog}) reflects
the well-known result that the matrix element of the operator $Q$
has the same leading logarithmic corrections in the static
theory (HQET) as the square of the decay constant, 
$f^2_{B_s}$. Taking $\mu_h=0.5,1,2\,$GeV results in 
$B_S(m_b)=0.80,0.88,0.94$. (The scale $\mu_h=0.5\,$GeV might 
already be too low for a perturbative evolution.)

The $b$-quark mass $m_b\approx 4.8\,$GeV is probably not large enough 
to make this estimate realistic, even if factorization held at 
the scale $\mu_h$. The logarithm $\ln m_b/\mu_h$
is not very large, so that other contributions like
non-logarithmic ${\cal O}(\alpha_s)$ terms 
which are omitted in (\ref{bhlog}),
can be expected to be numerically of the same order as the hybrid
logarithms that are retained, especially since summing hybrid 
logarithms amounts to a moderate $10\%$ effect (with $\mu_h=1\,$GeV). 
The one-loop matching of $Q$ on its counterpart(s) in Heavy 
Quark Effective Theory indeed exhibits sizeable cancellations 
between logarithms and constants, at least in the particular matching \
scheme considered in \cite{FLY91}. Furthermore, the QCD renormalization
between $m_b$ and $\mu_h$ in (\ref{bhlog}) is only valid at leading
order in HQET and neglects $1/m_b$ corrections in the matrix
elements, which is not consistent with our keeping of 
explicit $1/m_b$ corrections. On the other hand the $B$ factors 
are in principle calculable in
full QCD. In this case they will automatically include $1/m_b$
corrections as well as the hybrid logarithms, among further
important contributions. 
For these reasons we prefer to keep the expression for
$(\Delta\Gamma/\Gamma)_{B_s}$ in the form given in (\ref{dgres}) and do 
not include hybrid renormalization explicitly, with the
understanding that the bag factors will eventually be available
from lattice QCD. In our numerical analysis, we take the conservative, 
but perhaps too agnostic attitude that $B_S(m_b)$ could take any 
value between $0.7$ and $1.3$, keeping in mind (\ref{bhlog}) as 
a particular model estimate of $B$ and $B_S$.
The upper end of this range is motivated by the 
$N_c\to\infty$ limit, in which $B_S=6/5$.

\section{$\Delta\Gamma_{B_s}$ -- Phenomenology}
\label{pheno}

\subsection{Numerical Analysis of $(\Delta\Gamma/\Gamma)_{B_s}$}
\label{numeric}

\begin{table}[t]
\addtolength{\arraycolsep}{0.2cm}
\renewcommand{\arraystretch}{1.3}
$$
\begin{array}{|c|c||c|c|c|c|}
\hline
m_b/\mbox{GeV} & \mu & a & b & c & 
(\Delta\Gamma/\Gamma)_{B_s} \\ 
\hline\hline
4.8 & m_b & 0.009 & 0.211 & -0.065 & 0.155 \\ \hline
4.6 & m_b & 0.015 & 0.239 & -0.096 & 0.158 \\ \hline
5.0 & m_b & 0.004 & 0.187 & -0.039 & 0.151 \\ \hline
4.8 & 2 m_b & 0.017 & 0.181 & -0.058 & 0.140 \\ \hline
4.8 & m_b/2 & 0.006 & 0.251 & -0.076 & 0.181 \\ \hline
\end{array}
$$
\caption{\label{table1}
Dependence of $a$, $b$ and $c$ on the $b$-quark mass and renormalization 
scale for fixed values of all other short-distance parameters. The last 
column gives $(\Delta\Gamma/\Gamma)_{B_s}$ for $B=B_S=1$ (at 
the given scale $\mu$), $f_{B_s}=210\,$MeV.}
\end{table}

We first turn to a numerical analysis and discussion of  
$(\Delta\Gamma/\Gamma)_{B_s}$ based on (\ref{dgres}). It is 
useful to separate the dependence on the long-distance parameters 
$f_{B_s}$, $B$ and $B_S$ and write $(\Delta\Gamma/\Gamma)_{B_s}$ as 

\begin{equation}
\left(\frac{\Delta\Gamma}{\Gamma}\right)_{B_s} = 
\Big[a B + b B_S + c\Big]\left(\frac{f_{B_s}}{210\,\mbox{MeV}}\right)^2,
\end{equation}

\noindent where $c$ incorporates the explicit $1/m_b$ corrections. 
To estimate the sensitivity of $(\Delta\Gamma/\Gamma)_{B_s}$ on 
the short-distance input parameters, we keep the following parameters 
fixed: $m_b-m_c=3.4\,$GeV, $m_s=200\,$MeV, $\Lambda^{(5)}_{LO}=200\,
$MeV. In addition $M_{B_s}=5.37\,$GeV and the semileptonic branching 
ratio is $B(B_s\to X e\nu)=10.4\%$. Then $a$, $b$ and $c$ depend only 
on $m_b$ and the renormalization scale $\mu$. For some values of 
$m_b$ and $\mu$, the coefficients $a$, $b$, $c$ are listed in 
Table~\ref{table1}. For a central choice of parameters, which we take 
as $m_b=4.8\,$GeV, $\mu=m_b$, $B=B_S=1$ and $f_{B_s}=210\,$MeV, we 
obtain

\begin{equation}\label{dgnum1}
\left(\frac{\Delta\Gamma}{\Gamma}\right)_{B_s} = 0.220 - 0.065 = 0.155,
\end{equation}

\noindent where the leading term and the $1/m_b$ correction are
separately displayed. As seen from the Table, the dependence on $m_b$ 
is weak, but $(\Delta\Gamma/\Gamma)_{B_s}$ increases by almost $20\%$ 
when the renormalization scale is lowered to $m_b/2$, at fixed $B$ and 
$B_S$. These dependences are not specific to the values $B=B_S=1$. The 
weak $m_b$ dependence is a somewhat accidental consequence of using the 
semileptonic branching ratio to eliminate $V_{cb}$. If instead we 
normalize to $\Gamma_{B_s}^{-1}=1.54\,$ps and take $V_{cb}=0.04$, 
$(\Delta \Gamma/\Gamma)_{B_s}$ would vary from $0.143$ to $0.166$ under 
the same variation of $m_b$ as in the Table. Let us also add the 
following more general observations:

(i) The theoretical expression for $\Delta\Gamma_{B_s}$ in
(\ref{dgres}) predicts the sign of this quantity, which a
priori could have either value. $\Delta\Gamma_{B_s}$ is positive
and implies a larger decay rate for the CP even (lighter) state
\cite{VUKS,ALE93} 
(see the conventions in the Introduction). The typical magnitude of 
$(\Delta\Gamma/\Gamma)_{B_s}$ to leading order in the heavy quark 
expansion is about 0.2, larger than other width differences among 
bottom hadrons with the possible exception of the case of $\Lambda_b$ 
(depending on whether theory or present experiments turn out to be right 
on $\Lambda_b$).

(ii) The explicit $1/m_b$ corrections are numerically important 
and vary strongly with $m_b$. For our central parameter choice 
they reduce the leading order prediction by about $30\%$. 
Essentially all the various $1/m_b$ correction terms add with 
the same sign and make the result somewhat larger than the 
natural size of the corrections,
$\Lambda_{QCD}/m_b\approx (M_{B_s}-m_b-m_s)/m_b\approx 8\%$
and $m_s/m_b\approx 4\%$.

(iii) The contribution from the scalar operator $Q_S$ by far 
dominates over the contribution from $Q$, because there is a 
strong cancellation between terms of different sign in the Wilson
coefficient of the latter operator. This has important 
implications for $(\Delta M/\Delta\Gamma)_{B_s}$, which we 
discuss below, because hadronic uncertainties cancel only partially 
in the ratio $B/B_S$. 

(iv) If $B_S=1.3$, a $(\Delta\Gamma/\Gamma)_{B_s}$ of as much as 
$0.25$ is not excluded, although this appears unlikely. 
On the other hand, 
if $B_S<1$, as suggested by the estimate from hybrid logarithms, 
and if $f_{B_s}$ turns out to be merely 
$180\,$MeV, $(\Delta\Gamma/\Gamma)_{B_s}$ could be as small as 
$0.07$, making its experimental detection more difficult.

This discussion shows that to resolve the theoretical uncertainties, 
a reliable calculation of $B_S$ is mandatory. Further improvement 
then requires a full next-to-leading order calculation of short-distance 
corrections.

\subsection{Upper Limit on $\Delta\Gamma_{B_s}$}
\label{limit}

Since the $b\to c\bar cs$ transition is the dominant contributor to
$(\Delta\Gamma )_{B_s}$, one obtains the upper bound \cite{DUN1,atwood}

\begin{equation}\label{dglim}
\left(\frac{|\Delta\Gamma |}{\Gamma}\right)_{B_s} \leq 2 
\;B(b\rightarrow c\bar cs )_{B_s}\;.
\end{equation}

\noindent It can be readily understood by considering the limit in
which only $b\to c\bar cs$ 
transitions were generated by the effective Hamiltonian.
Eq. (\ref{dglim}) then follows from the requirement that the decay
rates be non-negative, 
$\Gamma_\pm =\Gamma(b\to c\bar cs)\pm\Delta\Gamma/2\geq 0$.
$B(b\rightarrow c\bar cs)_{B_s}$ 
denotes the fraction of $B_s$-meson decays 
governed by the $b\rightarrow c\bar cs$ transitions
in the absence of mixing. CLEO \cite{moriond} recently confirmed
our prediction \cite{bdy} of a significant `wrong' charm yield in 
$B$ decays, thereby completing the first direct measurement of

\begin{equation}
B(b\rightarrow c\bar cs^\prime ) \approx B(b\rightarrow \bar c )= 
0.227 \pm 0.035,
\end{equation}

\noindent where $B(b\rightarrow \bar c )$ is the average 
number of $\bar{c}$ produced per $b$ decay.
The Cabibbo allowed transition is 

\begin{equation}
B(b\rightarrow c\bar cs )=|V_{cs}|^2 \;B(b\rightarrow c\bar
cs^\prime )= 0.22 \pm 0.03,
\end{equation}

\noindent Assuming $B(b\to c\bar cs)_{B_s}\approx B(b\to c\bar cs)$ 
then yields the upper limit

\begin{equation}
\left(\frac{|\Delta\Gamma|}{\Gamma}\right)_{B_s} \leq 0.44 \pm 0.06 \;.
\end{equation}

\noindent Within the heavy quark expansion, 
$(|\Delta\Gamma|)/\Gamma)_{B_s}$ is suppressed by $m_b^{-3}$
relative to spectator branching ratios, 
such as $B(b\rightarrow c\bar cs)$.
{}From this point of view a bound like (\ref{dglim}) might appear
trivial. However, the virtue of relation (\ref{dglim}) is its very
general validity. It would hold even if a heavy quark expansion
were not applicable to the underlying process.

\subsection{Measuring $\Delta\Gamma_{B_s}$}
\label{measure}

We hope to have convinced the reader about the importance of an accurate
measurement of $\Delta\Gamma$.
One method is to substitute $\Gamma_{B_d}$ for the average $B_s$ 
width $\Gamma_{B_s}$ and to extract $\Delta\Gamma_{B_s}$ from the 
time-dependences of untagged flavor specific $B_s$ data samples 
\cite{DUN1}. Time-dependent studies of angular distributions of untagged
$\stackrel{(-)}{B}_s \rightarrow J/\psi\phi$ decays allow the extraction 
of $\Gamma_L$, and also of $\Gamma_H$ if the CP-odd component 
is non-negligible \cite{DDF,ddlr}.
These and other methods using decay length distributions of fully 
reconstructed $B_s$ mesons
are at present statistics limited~\cite{DUN1,DDF,ddlr}.

As an illustration one may consider the measurement of

\begin{equation}\label{tbsj}
\tau(B_s\to J/\psi \phi)=1.34^{+0.23}_{-0.19}\pm 0.05\,\mbox{ps}
\end{equation}

\noindent recently obtained by the CDF collaboration from a single 
lifetime fit of their $\stackrel{(-)}{B}_s \rightarrow J/\psi\phi$  
data sample \cite{MES}. Next we can write

\begin{equation}\label{glin}
1/\Gamma_L\leq\tau(B_s\to J/\psi\phi),
\end{equation}

\noindent which holds only as an inequality, 
because $B_s\to J/\psi\phi$ is not necessarily a pure CP-even final state.
The world average $B_d$ lifetime \cite{SCH}

\begin{equation}\label{tbdw}
\tau_{B_d}=1.54\pm 0.04\,\mbox{ps}
\end{equation}

\noindent together with the result of section \ref{bsbd}, informs us about 
the inverse of the average $B_s$ width $1/\Gamma_{B_s}=\tau_{B_d}$.
We then use

\begin{equation}\label{dggl}
\frac{\Delta\Gamma}{\Gamma}=
 2\left(\frac{\Gamma_L}{\Gamma}-1\right)
\end{equation}

\noindent and obtain

\begin{equation}
\left(\frac{\Delta\Gamma}{\Gamma}\right)_{B_s}\geq 0.3\pm 0.4,
\end{equation} 

\noindent which is still inconclusive, but can serve to indicate the present
status. 

Just establishing a non-vanishing difference in decay length 
distributions for partially reconstructed $B_s$ mesons in comparison 
to the other $B$ mesons would constitute progress.
The ideal inclusive $b$-hadron data sample should have large statistics 
and be highly enriched in $B_s$ decay products originating predominanty 
from a single mass eigenstate $B_L$ (or $B_H$). The last requirement 
maximizes differentiation between $B_s$ and other $B$-mesons.  
The $\phi\phi X$ final state serves as an example \cite{eigen}.
The probable decay chain is
$B_s\rightarrow D^+_s D^-_s X,$  which is
dominantly CP even \cite{ALE93}.
Both $D_s$'s then decay into $\phi$'s.
While $D_s$ is seen significantly in $\phi$'s, 
the $D^+$ is seen in $\phi$'s by
about a factor of 10 less and the $D^0$ even 
less than that \cite{distinguish}.
The background due to $B$-meson decays is thus controllable and further
suppressed because $B$'s prefer to be seen as $D^0$ over $D^+$ by a ratio of
2.7~\cite{yamamoto}.
If sufficient statistics is available, the $D^\pm_s \phi X$ sample would be
even better.

The inclusive $B_s\rightarrow \phi\ell^+ X$ sample with a high 
$P_{T,re\ell}$ lepton, is flavor specific. 
Its time dependence is governed by the sum of two exponentials,
${\rm exp}\left(-\Gamma_L t\right)+{\rm exp}\left(-\Gamma_H t\right)$.
Theory predicts $(\Gamma_L+\Gamma_H)/2 = 1/\tau_{B_d}$, 
but the observation of the two
exponents requires precise decay length and boost information, 
whose accuracy increases the more fully the $B_s$ is reconstructed.

The less reconstructed the $B_s$ data sample, the more important it is 
to have a mono-energetic source of $B_s$ mesons. Thus the 
more inclusive techniques tend to be more 
useful for $e^+ e^- \rightarrow Z^0$ 
experiments than at hadron accelerators.
Of course, fully reconstructed $B_s$ data samples allow clean 
measurements of $\Delta\Gamma_{B_s}$.

\subsection{$B_s-\bar B_s$ Mixing and CKM Elements}
\label{bbmix}

The traditional methods for observation of CP violation
and the extraction of CKM phases require to resolve the
rapid $\Delta M_{B_s} t$ oscillations of tagged $B_s$ data
samples \cite{tagged}.
Current vertexing technology allows to resolve such oscillations
for $\Delta M_{B_s} 
\;\raisebox{-.4ex}{\rlap{$\sim$}} \raisebox{.4ex}{$<$}\; 10\,$ps${}^{-1}$.
Thus the recent lower limit from the ALEPH
collaboration \cite{aleph}

\begin{equation}\label{dm66}
\Delta M_{B_s} > 6.6\,\mbox{ps}^{-1}\qquad (95\%\ \mbox{C.L.})
\end{equation}

\noindent is significant. It may indicate the need to develop new methods
capable of higher resolving power.
Reliable predictions of $\Delta M_{B_s}$ are therefore important
in order to plan future $B_s$ experiments, in particular if only
lower limits will be available with current vertex techniques.
The most straightforward method makes use of \cite{BBL}

\begin{equation}
\Delta M_{B_s} =\frac{G^2_F \;M^2_W}{6\pi^2}\eta_B \;\;S_0(x_t ) 
M_{B_s} B_{B_s} f^2_{B_s} |V_{ts}|^2, 
\end{equation}

\noindent where $x_t=m^2_t/M^2_W$.
The current relative uncertainty is about 50\% and is dominated
by the uncertainty in $B_{B_s}$ ($\pm 30\%$),
$f^2_{B_s}$ ($\pm 40\%$), $|V_{ts}|^2$ ($\pm 15\%$) and
$S_0(x_t )$ ($\pm 8\%$). The fractional uncertainty
on $\Delta M_{B_s}$ can be expected to decrease to $\sim 15\%$
by the year 2002, 
anticipating improvements in the accuracy of  the relevant
parameters $B_{B_s}$ ($\pm 10\%$),
$f^2_{B_s}$ ($\pm 5\%$), $|V_{ts}|^2$ ($\pm 5\%$) and
$S_0(x_t )$ ($\pm 3\%$).

A variant of this method uses the experimental value for
$\Delta M_{B_d}$ and the ratio

\begin{equation}
\frac{(\Delta M)_{B_s}}{(\Delta M)_{B_d}} = \frac{M_{B_s}}{M_{B_d}}
\;\;\frac{B_{B_s}\;f^2_{B_s}}{B_{B_d} \;f^2_{B_d}} \;\;\left |
\frac{V_{ts}}{V_{td}}\right |^2
\end{equation}

\noindent to predict $\Delta M_{B_s}$. This approach will be useful only
if the CKM ratio $|V_{ts}/V_{td}|^2$ is accurately known.

If the first observation of $B_s-\bar B_s$ mixing is a
nonvanishing $\Delta\Gamma_{B_s}$ rather than $\Delta M_{B_s}$,
then a complementary method to predict $\Delta M_{B_s}$
opens up, based on the quantity (see (\ref{dgres})) 

\begin{eqnarray}\label{dgdm}
\left(\frac{\Delta\Gamma}{\Delta M}\right)_{B_s} &=&
\frac{\pi}{2} \;\frac{m^2_b}{M^2_W} \;
\left|\frac{V_{cb}V_{cs}}{V_{ts}V_{tb}}\right|^2 \;
\frac{\sqrt{1-4z}}{\eta_B S_0(x_t)} \cdot \nonumber \\
&& \cdot \,\Bigg[\left(2(1-z)K_1+(1-4z)K_2\right)
\left(1+\frac{1}{N_c}\right) \\
&&\hspace*{0.3cm}+\,(1+2z)\left(K_2-K_1\right)
 \frac{M^2_{B_s}}{(m_b+m_s)^2}
 \left(2-\frac{1}{N_c}\right)\frac{B_S}{B}\,+\, \delta_{1/m}\Bigg]. 
\nonumber
\end{eqnarray}

\noindent This result is valid to next-to-leading order in the
$1/m_b$ expansion and to leading logarithmic accuracy in QCD.
We have again used factorization for the subleading $1/m_b$
corrections. Note that with the bag parameter $B$ as defined in
(\ref{meqs}), the appropriate QCD correction factor $\eta_B$ is
identical to $C_+(\mu)$ from (\ref{c21pm}) in the leading
logarithmic approximation.

In the ratio $\Delta\Gamma/\Delta M$ the decay constant cancels
and the CKM uncertainty is almost completely removed since
\begin{equation}\label{vctbs}
\left|\frac{V_{cb}V_{cs}}{V_{ts}V_{tb}}\right|^2= 1\pm 0.03.
\end{equation}
At present the accuracy of $\Delta\Gamma/\Delta M$ is still
rather poor, $\Delta\Gamma/\Delta M=(5.6\pm 2.6)\cdot 10^{-3}$. 
The breakdown of errors is as follows: $\pm 2.3$ from varying 
$B_S/B$ between $0.7$ and $1.3$, ${}^{+1.1}_{-0.7}$ from varying 
$\mu$ between $m_b/2$ and $2 m_b$, $\pm 0.4$ from 
$m_b=4.8\pm 0.2\,$GeV and $\pm 0.4$ from $m_t=176\pm 9\,$GeV. 
The dominant uncertainty is due to $B_S/B$, which has never
been studied before. It is conceivable that a lattice
study could actually calculate $B_S/B$ more
accurately than the bag parameters themselves, because some 
systematic uncertainties may be expected to cancel in the ratio.
The quantity $\Delta\Gamma/\Delta M$ might thus be calculable
rather precisely in the future and $\Delta M_{B_s}$ could then
be estimated from the observed $\Delta\Gamma_{B_s}$.
In conjunction with $\Delta M_{B_d}$ this would provide an
alternative way of determining the CKM ratio $|V_{ts}/V_{td}|$,
especially if the latter is around its largest currently
allowed value \cite{BP}.
The width difference, and hence its observability increases the
larger $|V_{ts}|\approx|V_{cb}|$ becomes. In contrast, the
ratio 
$\Gamma(B\to K^*\gamma)/\Gamma(B\to\{\varrho,\omega\}\gamma)$
is best suited for extracting small $|V_{ts}/V_{td}|$ ratios,
provided the long distance effects can be sufficiently well
understood \cite{rhogamma}.

These approaches could complement other methods to determine
$|V_{td}/V_{ts}|$. Such additional possibilities would be
to relate $|V_{ts}|$ to the accurate $|V_{cb}|$ measurements
and to obtain $|V_{td}|$ from $\Delta M_{B_d}$, CKM unitarity
constraints \cite{NIR2}, 
and in particular $B(K^+\to\pi^+\nu\bar\nu)$ \cite{BBL,BB}, 
which has the unique advantage of being
exceptionally clean from a theoretical point of view.

\section{The $B_s-B_d$ Width Difference}
\label{bsbd}

The ratio of the $B_s$ and $B_d$ meson decay widths 
$\Gamma_{B_s}/\Gamma_{B_d}$ is expected to be very close
to unity \cite{BIG,NES}. Deviations arise predominantly from 
$SU(3)$ breaking
effects in already small corrections to the leading spectator
decay of the bottom quark. In the following we will discuss
the mechanisms that differentiate between $\Gamma_{B_s}$
and $\Gamma_{B_d}$ and estimate their numerical importance.
The decay rate of $B_d$, $B_s$ mesons has the general form
($q=d$, $s$)

\begin{equation}\label{gabq}
\Gamma_{B_q}=\Gamma_0+\Delta\Gamma^{(q)}_{kin}+
\Delta\Gamma^{(q)}_{mag}+\Delta\Gamma^{(q)}_{WA} .
\end{equation}

\noindent Here $\Gamma_0$ denotes the leading, universal free $b$-quark
decay rate, $\Delta\Gamma_{kin}$ is the time dilatation
correction, $\Delta\Gamma_{mag}$ the contribution from the
chromomagnetic interaction of the heavy quark spin, and
$\Delta\Gamma_{WA}$ describes the weak annihilation of
$\bar b$ with $q$. $\Delta\Gamma_{kin}$ and $\Delta\Gamma_{mag}$
are of the order ${\cal O}(1/m^2_b)$ relative to $\Gamma_0$
and $\Delta\Gamma_{WA}$ enters at order ${\cal O}(1/m^3_b)$.
Higher orders have been neglected in (\ref{gabq}).
There is no linear correction in $1/m_b$ \cite{BIG}.
Through order ${\cal O}(1/m^3_b)$ one may thus write

\noindent \begin{equation}\label{gbsd}
\frac{\Gamma_{B_s}}{\Gamma_{B_d}}=1+
\frac{\Delta\Gamma^{(s)}_{kin}-\Delta\Gamma^{(d)}_{kin}}{\Gamma}+
\frac{\Delta\Gamma^{(s)}_{mag}-\Delta\Gamma^{(d)}_{mag}}{\Gamma}+
\frac{\Delta\Gamma^{(s)}_{WA}-\Delta\Gamma^{(d)}_{WA}}{\Gamma} .
\end{equation}

\noindent We will now discuss the three different corrections which 
contribute to $\Gamma_{B_s}/\Gamma_{B_d}-1$ in turn.

The first two can be related to meson mass differences.
For this purpose we define

\begin{equation}\label{mavg}
\overline{M}_H=\frac{1}{4}(M_H+3 M_{H^*}) ,
\end{equation}

\noindent where $M_H$ and $M_{H^*}$ are the masses of a pseudoscalar
heavy-light meson $H$ (${}^1S_0$) and of its vector meson
partner $H^*$ (${}^3S_1$). In the weighted average $\overline{M}_H$
the spin splitting contribution cancels in the HQET mass
formula which then takes the form ($Q=b$, $c$)

\begin{equation}\label{mhqet}
\overline{M}_{H_q}=m_Q+\bar{\Lambda}_q+
\frac{\langle\vec p^2\rangle_q}{2m_Q}+
{\cal O}\left(\frac{\Lambda^3_{QCD}}{m^2_Q}\right) .
\end{equation}

\noindent Here $\langle\vec p^2\rangle_q$ is the average momentum
squared of the heavy quark inside the meson and $\bar\Lambda_q$ may
be viewed as the constituent mass of the light degrees of
freedom. Both quantities depend on the light quark flavor $q$
but are independent of the heavy quark mass.
Combining (\ref{mhqet}) for the cases of $D_s$, $D^+$, $B_s$
and $B_d$ and recalling that 
$\Delta\Gamma^{(q)}_{kin}/\Gamma=-\langle\vec p^2\rangle_q/(2m_b^2)$
one finds

\begin{equation}\label{dgkin}
\frac{\Delta\Gamma^{(s)}_{kin}-\Delta\Gamma^{(d)}_{kin}}{\Gamma}=
-\frac{m_c/m_b}{m_b-m_c}\left[\overline{M}_{D_s}-\overline{M}_{D^+}-
(\overline{M}_{B_s}-\overline{M}_{B_d})\right]
\approx -(3\pm 6)\cdot 10^{-4} .
\end{equation}

\noindent All required 
meson masses can be obtained from \cite{PDG}, except for
$M_{B^*_s}$. In this case we use the heavy quark symmetry relation

\begin{equation}\label{mbss}
M_{B^*_s}-M_{B_s}=
\frac{M_{D^*_s}-M_{D_s}}{M_{D^{*+}}-M_{D^+}}(M_{B^*_d}-M_{B_d})
=(46\pm 1)\ \mbox{MeV}
\end{equation}

\noindent to find $M_{B^*_s}=(5421\pm 6)\,$MeV. This expectation is
in accordance with direct measurements of the
$B^*_s\to B_s\gamma$ transition, which yield
$M_{B^*_s}-M_{B_s}=(47.0\pm 2.6)\,$MeV \cite{LEE}.
We see that the correction in (\ref{dgkin}) is exceedingly small.
This number, however, should probably not be taken at face value.
Given the smallness of the effect it is conceivable that terms
neglected in (\ref{mhqet}) could have an impact on the precise
estimate of (\ref{dgkin}). The typical size of such a correction
would be (here we use $\Lambda_{QCD}=0.3$\ GeV)

\begin{equation}\label{dgk2}
\left|\frac{\Delta\Gamma^{(s)}_{kin}-\Delta\Gamma^{(d)}_{kin}}{
\Gamma}\right|\approx\frac{m_c/m_b}{m_b-m_c}
\left[\frac{\Lambda^3_{QCD}}{m^2_c}\right]\approx 12\cdot 10^{-4} .
\end{equation}

\noindent At any rate, while (\ref{dgkin}) might not be a completely
accurate estimate of this correction, it seems safe to conclude
that the effect on $\Gamma_{B_s}/\Gamma_{B_d}$ due to
$\Delta\Gamma^{(s)}_{kin}-\Delta\Gamma^{(d)}_{kin}$ is well
below 1\% and thus negligible for all practical purposes.

Next, the chromomagnetic correction $\Delta\Gamma^{(q)}_{mag}$
can be related to the spin splitting in S-wave $B$ mesons
and is proportional to $M_{B^*_q}-M_{B_q}$. Hence we may write

\begin{equation}\label{dgmag}
\frac{\Delta\Gamma^{(s)}_{mag}-\Delta\Gamma^{(d)}_{mag}}{\Gamma}=
\frac{\Delta\Gamma^{(d)}_{mag}}{\Gamma}
\frac{M_{B^*_s}-M_{B_s}-(M_{B^*_d}-M_{B_d})}{M_{B^*_d}-M_{B_d}}
\approx -(3\pm 8)\cdot 10^{-4} .
\end{equation}

\noindent The quantity 
$\Delta\Gamma^{(d)}_{mag}/\Gamma$ is known \cite{BIG}
and can be calculated to be $-0.012$.
Using $M_{B^*_d}-M_{B_d}=(46.0\pm 0.6)$\ MeV \cite{PDG}
and $M_{B^*_s}-M_{B_s}=(47.0\pm 2.6)$\ MeV \cite{LEE}
one finds the numerical estimate quoted in (\ref{dgmag}). Clearly
this effect on the $B_s -B_d$ lifetime difference is negligible
as well.

Finally, we turn to the corrections  due to weak annihilation.
These contributions arise from the annihilation reactions
$\bar bs\to\bar cc$ and $\bar bd\to\bar cu$ in the case of
a $B_s$ and a $B_d$ meson, respectively.
Neglecting Cabibbo suppressed modes and penguin contributions
they are readily calculated to be\footnote{
Our results are in agreement with the expressions recently
obtained in \cite{NES}.}

\begin{eqnarray}\label{dgwas}
\frac{\Delta\Gamma^{(s)}_{WA}}{\Gamma} &=&
16\pi^2 B(B\to Xe\nu)\,\frac{f^2_{B_s}}{m^2_b}\,V^2_{cs}\,
\frac{\sqrt{1-4z}}{g(z)\,\tilde{\eta}_{QCD}}\cdot  \\
&&\cdot\left[-(1-z)\left(K_1 B^{(s)}_1+
\frac{1}{N_c}K_2 B^{(s)}_2\right)+
(1+2z)\left(K_1 B^{(s)}_3+\frac{1}{N_c}K_2 B^{(s)}_4\right)\right], 
\nonumber
\end{eqnarray}

\begin{eqnarray}\label{dgwad}
\frac{\Delta\Gamma^{(d)}_{WA}}{\Gamma} &=&
16\pi^2 B(B\to Xe\nu)\,\frac{f^2_{B_d}}{m^2_b}\,V^2_{ud}\,
\frac{(1-z)^2}{g(z)\,\tilde{\eta}_{QCD}}\cdot  \\
&&\cdot\left[-(1+\frac{z}{2})
\left(K_1 B^{(d)}_1+\frac{1}{N_c}K_2 B^{(d)}_2\right)+
(1+2z)\left(K_1 B^{(d)}_3+\frac{1}{N_c}K_2 B^{(d)}_4\right)\right] . 
\nonumber
\end{eqnarray}

\noindent Here we have again used (\ref{gbsl}) to eliminate the $V_{cb}$
dependence. The leading log QCD coefficients $K_{1,2}$ are
defined in (\ref{k1k2}). The bag factors $B^{(q)}_i$
parametrize the following matrix elements

\begin{eqnarray}\label{b1234}
\langle B_q|(\bar b_iq_i)_{V-A}(\bar q_jb_j)_{V-A}|B_q\rangle
&=& f^2_{B_q} m^2_b B^{(q)}_1 \nonumber \\
\langle B_q|(\bar b_iq_j)_{V-A}(\bar q_jb_i)_{V-A}|B_q\rangle
&=& \frac{1}{N_c} f^2_{B_q} m^2_b B^{(q)}_2  \nonumber \\
\langle B_q|(\bar b_iq_i)_{S-P}(\bar q_jb_j)_{S+P}|B_q\rangle
&=& f^2_{B_q} m^2_b B^{(q)}_3 \nonumber \\
\langle B_q|(\bar b_iq_j)_{S-P}(\bar q_jb_i)_{S+P}|B_q\rangle
&=& \frac{1}{N_c} f^2_{B_q} m^2_b B^{(q)}_4 ,
\end{eqnarray}

\noindent where we have assumed $M_{B_q}\approx m_b$.

Using the strict factorization estimate $B^{(q)}_i\equiv 1$ would
yield the following result (taking $f_{B_d}\approx f_{B_s}$ and 
expanding in $z\approx 0.1$)

\begin{equation}\label{dgwa1}
\left[
\frac{\Delta\Gamma^{(s)}_{WA}-\Delta\Gamma^{(d)}_{WA}}{\Gamma}
\right]_{fact.}\simeq
24\pi^2 B(B\to Xe\nu)\,\frac{f^2_B}{m^2_b}\,
\frac{1-2z}{g(z)\,\tilde{\eta}_{QCD}}
\, z\,\left(K_1+\frac{1}{N_c}K_2\right) .
\end{equation}

\noindent Note that, in `vacuum insertion', this expression coincides
with $\Delta\Gamma^{(d)}_{WA}/\Gamma$ while
$\Delta\Gamma^{(s)}_{WA}/\Gamma$ is twice as large.
For our central parameter set, eq.~(\ref{dgwa1}) amounts to
$2\cdot 10^{-4}$. The extreme smallness of this number
is the result of two effects. The first is helicity suppression, 
manifesting itself in the factor of $z=m^2_c/m^2_b$ in (\ref{dgwa1}).
Secondly, a further suppression comes from a -- somewhat accidental --
cancellation between QCD coefficients in 
$K_1+K_2/3\approx -0.39+0.42=0.03$. It is important to
realize that both features are a consequence of the factorization
assumption. Even with small deviations from factorization the factor 
$z (K_1+K_2/N_c)$ would be substituted by a number 
almost one hundred times 
larger. To get an idea of the typical order of magnitude, 
we approximate (\ref{dgwas}) and (\ref{dgwad})
to

\begin{eqnarray}\label{dgwa2}
\frac{\Delta\Gamma^{(q)}_{WA}}{\Gamma} &=&
16\pi^2 B(B\to Xe\nu)\,\frac{f^2_{B_q}}{m^2_b}\,V^2_{ud}\,
\frac{1-2z}{g(z)\,\tilde{\eta}_{QCD}}\cdot  \\
&&\cdot\left[
K_1 (B^{(q)}_3-B^{(q)}_1)+\frac{1}{N_c}K_2(B^{(q)}_4-B^{(q)}_2) +
 {\cal O}(z) \right], 
\nonumber
\end{eqnarray}

\noindent where we have used $\sqrt{1-4z}\approx (1-z)^2\approx 1-2z$
and neglected small helicity suppressed contributions proportional to $z$
in the square brackets. Taking $-K_1\approx K_2/3\approx 0.4$
and $|B^{(q)}_3-B^{(q)}_1|,|B^{(q)}_4-B^{(q)}_4|<0.6$, the modulus of 
the term in square brackets is $0.5$ or less,
which yields $\Delta\Gamma^{(q)}_{WA}/\Gamma\leq 0.023$.
Assuming $40\%$ of $SU(3)$ breaking then gives

\begin{equation}\label{dgwa3}
\left|\frac{\Delta\Gamma^{(s)}_{WA}-\Delta\Gamma^{(d)}_{WA}}{\Gamma}
\right|\leq 0.9\% .
\end{equation}

\noindent Although with extreme variations, allowing also $|K_1|$ 
and $|K_2/3|$ to differ (for example by choosing a renormalization 
scale $\mu$ different from $m_b$), this difference could be up to 
$2.5\%$, it is more likely that the correction (\ref{dgwa3}) will
actually be much smaller due to various possible cancellations
in (\ref{dgwa2}) and because $40\%$ is probably an overestimate
of the magnitude of $SU(3)$ breaking. Furthermore, from previous
experience with lattice calculations of bag parameters in the
$B$ meson system it seems likely that the $B^{(q)}_i$ will
not differ too dramatically from one, so that (\ref{dgwa3}),
although admittedly somewhat crude, is probably on the safe side.

Summarizing the discussion of the various contributions to
(\ref{gbsd}) we conclude that, most likely, the ratio of rates of
$B_s$ and $B_d$ mesons should differ from unity by no more
than one percent

\begin{equation}\label{gbsd1}
\left|\frac{\Gamma_{B_s}}{\Gamma_{B_d}}-1\right|< 1\% .
\end{equation}

\section{Summary}
\label{summary}

In this paper we have analyzed the theoretical prediction
for $\Delta\Gamma_{B_s}$ within the framework of the heavy
quark expansion.
We have calculated the explicit next-to-leading ${\cal O}(1/m_b)$
corrections in the operator product expansion for the
transition matrix element. In addition to the two leading 
dimension-six operators, five new operators of dimension
seven appear at this level. The matrix elements of the latter
operators were evaluated using factorization, which should give
a fair estimate of these subleading corrections. 
Their effect on $\Delta\Gamma_{B_s}$, 
formally of order ${\cal O}(\Lambda_{QCD}/m_b)$ and
${\cal O}(m_s/m_b)$, turned out to be sizable numerically,
causing a $30\%$ reduction of the leading order prediction.

We performed a numerical investigation of 
$\Delta\Gamma_{B_s}$ with emphasis
on theoretical errors, which are presently dominated
by the uncertainties in hadronic matrix elements. 
These errors are still rather large and lead to a prediction
of $(\Delta\Gamma/\Gamma)_{B_s}=0.16^{+0.11}_{-0.09}$.
However, a systematic improvement of this result is possible,
in particular by progress in lattice QCD. In the future it would 
be desireable to measure on the lattice the S-P four-fermion operator 
along with the V-A operator that has received most attention 
in the past due to its connection with the mass difference. 
Eventually an accuracy
of $10\%$ for $\Delta\Gamma_{B_s}$ should be feasible when the 
next-to-leading analysis of short-distance corrections is also 
completed.

The effects of penguin operators and contributions
from CKM suppressed modes have also been considered. They were
shown to give only a few percent relative correction in
$(\Delta\Gamma/\Gamma)_{B_s}$ and are thus negligible in view
of the other uncertainties.

We further studied the $B_s-B_d$ lifetime difference and quantified 
the expectation $\tau_{B_s}\approx\tau_{B_d}$, estimating
$|\tau(B_s)/\tau(B_d)-1|<1\%$. This result is useful input for
experimental analyses of $\Delta\Gamma_{B_s}$.

To put our theoretical analysis into perspective, we have included 
a short discussion of the current experimental situation
concerning $\Delta\Gamma_{B_s}$. Using information on
$\tau(B_s\to J/\psi\phi)$ and $\tau(B_s)=\tau(B_d)$,
we have attempted a preliminary extraction of $\Delta\Gamma_{B_s}$,
obtaining $(\Delta\Gamma/\Gamma)_{B_s}\geq 0.3\pm 0.4$. This is still
inconclusive but can be improved by better statistics in the
future. We have also
proposed an alternative route towards a measurement of
$\Delta\Gamma_{B_s}$ that makes use of the 
$\phi\phi X$ and/or $D^\pm_s\phi X$ final states
in $B_s$ decay, which are expected to be dominantly CP even.
The present experimental information may be complemented by the
bound  
$(\Delta\Gamma/\Gamma)_{B_s}\leq 2 B(b\to c\bar cs)_{B_s}
\approx 0.44\pm 0.06$.
This bound is not very strong, but it has the advantage of being
valid independently of the heavy quark expansion and it is interesting
for principal reasons.

In addition we have briefly reviewed some phenomenological applications
that could be opened up by further progress on the experimental
as well as the theoretical side. These possibilities include
new methods to study CP violation, complementary information on
$\Delta M_{B_s}$ in case $\Delta\Gamma_{B_s}$ is measured first,
and alternative constraints on $|V_{td}/V_{ts}|$, especially
for small values of this ratio.
Finally, the theory of inclusive $B$ decays itself can be
expected to profit from a confrontation of the heavy quark expansion
for $\Delta\Gamma_{B_s}$ with experiment. 
In this respect $\Delta\Gamma_{B_s}$ provides an important special
case that directly probes ${\cal O}(1/m^3_b)$ contributions. 

As we have seen, the topic of $\Delta\Gamma_{B_s}$ touches upon
a rich variety of interesting physics issues and certainly merits
the continued efforts needed to address the problems that are
still unresolved.

\vspace*{0.7cm}

\noindent {\bf Acknowledgements.} 
We thank Andrzej Buras for encouragement and useful
conversations. Thanks are also due to Yuval Grossman,
Joe Incandela, Jonathan Lewis, Hans-G\"unther Moser
and Yossi Nir for discussions.
Fermilab is operated by Universities Research Association, Inc.,
under contract DE-AC02-76CHO3000 with the United States Department
of Energy.

\section*{Appendix A: Penguin Contributions to $\Delta\Gamma_{B_s}$}
\label{appa}

In the following we discuss the impact of penguin operators
on the width difference $\Delta\Gamma_{B_s}$.
We will work to leading logarithmic accuracy in QCD and include
the charm quark mass effects. For the purpose of this section
we shall neglect $1/m_b$ corrections, CKM suppressed modes and
light quark masses.

Taking gluonic penguin operators into account, the effective
hamiltonian in (\ref{heff}) is generalized to

\begin{equation}\label{hpeng}
{\cal H}_{eff}=\frac{G_F}{\sqrt{2}}V^*_{cb}V_{cs}
\sum^6_{r=1} C_r Q_r
\end{equation}

\noindent where

\begin{equation}\label{q1q2}
Q_1= (\bar b_is_i)_{V-A}(\bar c_jc_j)_{V-A}\qquad
Q_2= (\bar b_is_j)_{V-A}(\bar c_jc_i)_{V-A}
\end{equation}
\begin{equation}\label{q3q4}
Q_3= (\bar b_is_i)_{V-A}(\bar q_jq_j)_{V-A}\qquad
Q_4= (\bar b_is_j)_{V-A}(\bar q_jq_i)_{V-A}
\end{equation}
\begin{equation}\label{q5q6}
Q_5= (\bar b_is_i)_{V-A}(\bar q_jq_j)_{V+A}\qquad
Q_6= (\bar b_is_j)_{V-A}(\bar q_jq_i)_{V+A}.
\end{equation}

\noindent A summation over $q=u$, $d$, $s$, $c$ is implied.
$C_1,\ldots, C_6$ are the corresponding Wilson coefficient
functions. $C_{1,2}$ have already been given in (\ref{c21pm}).
For a recent review of this subject see \cite{BBL}, where further 
details may be
found. Using our standard parameter set with $\mu=m_b$
the numerical values are 

\begin{equation}\label{c16num}
(C_1, \ldots, C_6)=(-0.272,1.120,0.012,-0.028,0.008,-0.035).
\end{equation}

\noindent The calculation of the transition operator (\ref{tdef}) using
the extended operator basis is straightforward and leads to

\begin{eqnarray}\label{dgpeng}
&&\left(\frac{\Delta\Gamma}{\Gamma}\right)_{B_s} = 16\pi^2
B(B_s\to X e\nu)\frac{f^2_{B_s}M_{B_s}}{m^3_b} 
\frac{V^2_{cs}}{g(z)\,\tilde\eta_{QCD}}\ 
\Biggl\{\sqrt{1-4z} \cdot \nonumber \\
&&\cdot \left[
\left(2(1-z)(K_1+K_1'+K_1'')+(1-4z)(K_2+K_2'+K_2'')+6z(K_3'+K_3'')
    \right)\left(1+\frac{1}{N_c}\right)B+ \right. \nonumber\\
&&+ \left.(1+2z)(K_2+K_2'+K_2''-K_1-K_1'-K_1'')
 \frac{M^2_{B_s}}{(m_b+m_s)^2}
 \left(2-\frac{1}{N_c}\right)B_S\right] + \nonumber\\
&& +3\left[(2K_1''+K_2'')\left(1+\frac{1}{N_c}\right)B+
 (K_2''-K_1'')\frac{M^2_{B_s}}{(m_b+m_s)^2}
 \left(2-\frac{1}{N_c}\right)B_S\right]\Biggr\}.
\end{eqnarray}

\noindent $K_{1,2}$ are defined in (\ref{k1k2}) and the remaining
coefficients read

\begin{equation}\label{k12pr}
K_1'=2(N_c C_1 C_3+C_1 C_4+C_2 C_3)\qquad K_2'=2 C_2 C_4
\end{equation}
\begin{equation}\label{k3pr}
K_3'=2(N_c C_1 C_5+C_1 C_6+C_2 C_5+ C_2 C_6),
\end{equation}

\begin{equation}\label{k12dp}
K_1''=N_c C^2_3+N_c C^2_5+ 2 C_3 C_4+ 2 C_5 C_6\qquad K_2''= C^2_4+ C^2_6
\end{equation}
\begin{equation}\label{k3dp}
K_3''=2(N_c C_3 C_5+C_3 C_6+C_4 C_5+ C_4 C_6).
\end{equation}

\noindent These expressions represent the interference of penguin operators
with the leading operators $Q_{1,2}$ (coefficients $K_i'$) and
penguin-penguin contributions (coefficients $K_i''$).
Numerically they reduce $(\Delta\Gamma/\Gamma)_{B_s}$
by 0.0114, which is about 5\% of the result without penguins
$(\Delta\Gamma/\Gamma)_{B_s}=0.221$, neglecting $1/m_b$ 
corrections. Note that since
$C_3,\ldots, C_6$ are small, the effect of penguins is dominated
by the $K_i'$, while the $K_i''$ are negligible.

\section*{Appendix B: Cabibbo-suppressed Contributions 
to $\Delta\Gamma_{B_s}$}
\label{appb}

In this appendix we briefly consider the CKM suppressed
contributions to $\Delta\Gamma_{B_s}$. They arise from
$u\bar c$ ($\bar uc$) or $u\bar u$ intermediate states in
the diagram of Fig. \ref{fig1}.
For our estimate we include again QCD corrections in the leading
logarithmic approximation and keep charm quark mass effects.
We neglect $1/m_b$ corrections and the small impact of
penguin operators in the $u\bar u$ channel.

The contribution from $u\bar c$ and $\bar uc$ intermediate states
is then found to be

\begin{eqnarray}\label{dguc}
&&\left(\frac{\Delta\Gamma}{\Gamma}\right)_{B_s,uc} = 16\pi^2
B(B_s\to X e\nu)\frac{(1-z)^2}{g(z)\,\tilde\eta_{QCD}}
\frac{f^2_{B_s}M_{B_s}}{m^3_b} V^2_{cs} \cdot
  2 \,\mbox{Re}\,\frac{\lambda_u}{\lambda_c}\cdot \nonumber \\
&&\quad \cdot \left[\left((2+z)K_1+(1-z)K_2\right)
\left(1+\frac{1}{N_c}\right)B+ (1+2z)\left(K_2-K_1\right)
 \left(2-\frac{1}{N_c}\right)B_S \right],
\end{eqnarray}

\noindent where $\lambda_i=V^*_{ib}V_{is}$.
Compared to the leading, CKM allowed contribution with two
charm quarks in the intermediate state, expression
(\ref{dguc}) is suppressed by a factor

\begin{equation}\label{reluc}
2\,\mbox{Re}\,\frac{\lambda_u}{\lambda_c}=2\lambda^2 \varrho
\leq\pm 3\%
\end{equation}

\noindent Here we have used the Wolfenstein parametrization and the
result that $\varrho$ is restricted by $|\varrho|<0.3$
in the standard model \cite{BBL}. Since the difference
between (\ref{dguc}) and the CKM allowed contribution
(see (\ref{dgres})) due to the different charm quark mass
dependences turns out to be negligible numerically, 
relation (\ref{reluc}) determines essentially the relative 
importance of (\ref{dguc}) for $(\Delta\Gamma/\Gamma)_{B_s}$.
Note that the sign of (\ref{dguc}) is not yet fixed
because both positive and negative values are still allowed for
$\varrho$. Since $\varrho$ could be close to zero, the
CKM suppressed contribution (\ref{dguc}) might also be well below 
the $3\%$ given above. In any case, it can be safely neglected.

The contribution with two internal up quarks can
be obtained from (\ref{dguc}) by replacing
$2\,\mbox{Re}\,(\lambda_u/\lambda_c)\to \mbox{Re}\,(\lambda_u/\lambda_c)^2$
and setting $z\to 0$ everywhere except in the argument of $g(z)$.
Since $|\mbox{Re}\,(\lambda_u/\lambda_c)^2|$ can be estimated to be
smaller than $4\cdot 10^{-4}$, 
the resulting expression is still much more suppressed
than (\ref{dguc}) and therefore completely irrelevant.

\section*{Appendix C: Comment on $\tau_{B^+}/\tau_{B_d}$}
\label{bpbd}

Some of the issues in the calculation of lifetime differences
among $B_s$ and $B_d$ mesons that we have discussed in 
this paper are also relevant for the prediction of
$\tau_{B^+}/\tau_{B_d}$. We will therefore take the opportunity
to also have a brief look at the question of the $B^+-B_d$
lifetime difference.
In the literature this quantity has been estimated to be
\cite{BIG}

\begin{equation}\label{tbpd}
\frac{\tau_{B^+}}{\tau_{B_d}}\simeq 1+0.05\cdot
\frac{f^2_B}{(200\,{\rm MeV})^2},
\end{equation}

\noindent predicting the $B^+$ lifetime to exceed $\tau_{B_d}$ by
several percent. In the following we would like to re-examine
this estimate, emphasizing the theoretical uncertainties that are
involved in its derivation.
Assuming isospin symmetry, the mechanisms that produce a difference
in $\tau_{B^+}$ and $\tau_{B_d}$ first enter at the level of
dimension six operators, or equivalently at ${\cal O}(1/m^3_b)$,
in the heavy quark expansion \cite{BIG}. These effects are
weak annihilation for the $B_d$ and Pauli interference in the
case of $B^+$. As we have seen in section \ref{bsbd}, the
weak annihilation contribution to $\tau_{B_d}$ is very small
and we shall neglect it. In this approximation the difference
between $\tau_{B^+}$ and $\tau_{B_d}$ arises only through
Pauli interference and one may write

\begin{equation}\label{tbpi}
\frac{\tau_{B^+}}{\tau_{B_d}}=1+24\pi^2 B(B\to Xe\nu)\,
\frac{f^2_B}{m^2_b} V^2_{ud} \frac{(1-z)^2}{g(z)\,\tilde{\eta}_{QCD}}
\left[(C^2_--C^2_+)B^{(u)}_1-\frac{1}{N_c}
(C^2_++C^2_-)B^{(u)}_2\right],
\end{equation}

\noindent where

\begin{eqnarray}\label{bu12}
\langle B^+|(\bar b_iu_i)_{V-A}(\bar u_jb_j)_{V-A}|B^+\rangle
&=& f^2_{B} m^2_b B^{(u)}_1 \nonumber \\
\langle B^+|(\bar b_iu_j)_{V-A}(\bar u_jb_i)_{V-A}|B^+\rangle
&=& \frac{1}{N_c} f^2_{B} m^2_b B^{(u)}_2  
\end{eqnarray}

\noindent define the bag parameters $B^{(u)}_{1,2}$.
The Wilson coefficients $C_\pm$ have been given 
in (\ref{c21pm}).

With $m_b=4.8\,$GeV, $m_c=1.4\,$GeV, $\Lambda_{LO}=0.2\,$GeV
and taking $B^{(u)}_{1,2}=1$, $f_B=0.2$\,GeV, one finds
$\tau_{B^+}/\tau_{B_d}=1.02$, indicating a slightly longer
lifetime for $B^+$ than for $B_d$. This number can however not
be viewed as a very accurate prediction. In fact, the two
contributions proportional to $B^{(u)}_1$ and $B^{(u)}_2$
in (\ref{tbpi}) enter with different sign. This leads to a
partial cancellation that has the tendency to make the result
unstable. For instance, allowing the unphysical scale 
$\mu={\cal O}(m_b)$ in the coefficients $C_\pm$ to vary from 
$m_b/2$ to $2m_b$ gives a range of $1.00 - 1.06$ for the
$B^+$ to $B_d$ lifetime ratio. Switching off short distance
QCD corrections completely ($C_\pm\to 1$), the hierarchy of
lifetimes would even be reversed to $\tau_{B^+}/\tau_{B_d}=0.95$,
which is another aspect of the large sensitivity to QCD effects.
An alternative way of estimating the present uncertainty is to
allow a variation in the bag parameters (keeping $\mu=m_b$ fixed).
A range of $B^{(u)}_{1,2}=1.0\pm 0.3$ is certainly conceivable,
considering the uncertainties in the nonperturbative dynamics
and from the scale and scheme dependence in the 
long-distance to short-distance matching. Assuming this,
we obtain for $f_B=0.2$\,GeV, $\tau_{B^+}/\tau_{B_d}=1.02\pm 0.04$.
A combination of both variations, of scale and bag parameters, 
would even allow us to obtain a lifetime difference of up to $20\%$, 
$\tau_{B^+}/\tau_{B_d}\sim 1.2$. Although we consider this case highly 
unlikely the point to note is that a lifetime that large could be 
tolerated by QCD as well as equal
lifetimes, or even a marginally shorter lifetime for the $B^+$.
A decisive improvement of this situation could only be achieved
by a reliable lattice calculation of $B^{(u)}_{1,2}$ in
conjunction with a next-to-leading order computation of
short-distance QCD corrections to ensure a proper matching
in renormalization scheme and scale between Wilson coefficients
and hadronic matrix elements. Alternatively one could use the 
present measurement $\tau_{B^+}/\tau_{B_d}=1.06\pm 0.04$ \cite{SCH}
to constrain the bag parameters. At present such constraints 
appear to be of limited use, because of the large renormalization 
scale dependence of Pauli interference at leading order. 
Similar conclusions have been reached in the recent paper
by Neubert and Sachrajda \cite{NES}.

The authors of \cite{BIG} have modeled the bag parameters
in their estimate of $\tau_{B^+}/\tau_{B_d}$ by factorizing
at a low scale $\mu_h< m_b$ and explicitly including the
leading logarithms of HQET. This yields

\begin{equation}\label{buhl}
B^{(u)}_1(m_b)=\frac{8}{9}
\left[\frac{\alpha_s(m_b)}{\alpha_s(\mu_h)}\right]^{-3/50}+
\frac{1}{9} 
\left[\frac{\alpha_s(m_b)}{\alpha_s(\mu_h)}\right]^{12/25}
\qquad
B^{(u)}_2(m_b)=
\left[\frac{\alpha_s(m_b)}{\alpha_s(\mu_h)}\right]^{12/25} .
\end{equation}

\noindent Taking $\mu_h=1$\,GeV this gives $B^{(u)}_1(m_b)=1.01$,
$B^{(u)}_2(m_b)=0.72$ and $\tau_{B^+}/\tau_{B_d}=1.04$
(for $f_B=0.2$\,GeV), favoring $\tau_{B^+}>\tau_{B_d}$.
However, as discussed at the end of section \ref{basic}, the 
quantitative reliability of an estimate based on hybrid logarithms 
is not entirely clear.

\vfill\eject


\newpage

\begin{thebibliography}{99}
\bibitem{DUN1}
I. Dunietz, Phys. Rev. {\bf D52}, 3048 (1995)
\bibitem{BP}
T.E. Browder and S. Pakvasa, 
Phys. Rev. {\bf D52}, 3123 (1995)
\bibitem{BIG}
I. Bigi {\it et al.},
{\it in} B Decays, second edition, ed. S. Stone
(World Scientific, Singapore, 1994)
\bibitem{untag}
D. Atwood, I. Dunietz and A. Soni, FERMILAB-PUB-94/388-T,
in progress;
R. Fleischer and I. Dunietz, FERMILAB-PUB-96/079-T 
[hep-ph/9605220]; FERMILAB-PUB-96/080-T [hep-ph/9605221]
\bibitem{DDF}
A.S. Dighe, I. Dunietz and R. Fleischer, 
FERMILAB-PUB-96/081-T, in progress
\bibitem{HAG}
J.S. Hagelin, Nucl. Phys. {\bf B193}, 123 (1981) 
\bibitem{FLP}
E. Franco, M. Lusignoli and A. Pugliese, 
Nucl. Phys. {\bf B194}, 403 (1982)
\bibitem{CHA}
L.L. Chau, Phys. Rep. {\bf 95}, 1 (1983) 
\bibitem{BSS}
A.J. Buras, W. S\l ominski and H. Steger, 
Nucl. Phys. {\bf B245}, 369 (1984) 
\bibitem{VUKS}
M.B. Voloshin, N.G. Uraltsev, V.A. Khoze and M.A. Shifman, 
Sov. J. Nucl. Phys. {\bf 46}, 112 (1987) 
\bibitem{ALE93} 
R. Aleksan et al., Phys. Lett. {\bf B316}, 567 (1993)
\bibitem{GRO96} Y. Grossman, Weizmann Institute report WIS-96/13/Mar-PH 
[hep-ph/9603244]
\bibitem{PDG}
L. Montanet et al., Particle Data Group, Phys. Rev. {\bf D50}, 
1173 (1994)
\bibitem{BJW}
A.J. Buras, M. Jamin and P.H. Weisz, 
Nucl. Phys. {\bf B347}, 491 (1990) 
\bibitem{NOV} V.~Novikov, M.~Shifman, A.~Vainshtein and V.~Zakharov, 
Fortsch.~Phys. {\bf 32}, 585 (1984); In the context of heavy quark 
expansions, see also M.~Shifman, TASI lectures 1995 [hep-ph/9510377]
\bibitem{KIL93} W.~Kilian and T.~Mannel, Phys.~Lett. {\bf B301}, 
382 (1993)
\bibitem{NIR} Y.~Nir, Phys.~Lett. {\bf B221}, 184 (1989) 
\bibitem{VS}
M.B. Voloshin and M.A. Shifman, 
Sov. J. Nucl. Phys. {\bf 45}, 292 (1987) 
\bibitem{PW}
H. Politzer and M. Wise, Phys. Lett. {\bf B206}, 681 (1988); 
{\bf B208}, 504 (1988)
\bibitem{FLY91} J.~Flynn, O.~Hern\'{a}ndez and B.~Hill, Phys.~Rev. 
{\bf D43}, 3709 (1991)
\bibitem{atwood}
We are grateful to David Atwood for pointing out to us that the
right-hand-side must have a factor of 2, which is missing in \cite{DUN1}.
See also: D. Atwood, I. Dunietz and J. Incandela,
FERMILAB-PUB-95/402-T, in progress
\bibitem{moriond}
Y. Kwon, talk given at Moriond, March 1996
\bibitem{bdy}
G. Buchalla, I. Dunietz and H. Yamamoto, 
Phys. Lett. {\bf B364}, 188 (1995)
\bibitem{ddlr}
A.S. Dighe, I. Dunietz, H.J. Lipkin and J.L. Rosner,
Phys. Lett. {\bf B369}, 144 (1996)
\bibitem{MES}
E. Meschi, FERMILAB-CONF-96/013-E
\bibitem{SCH}
We thank Slawek Tkaczyk, who averaged the available CDF measurements
with the most recent LEP average as reported in
M.-H. Schune, talk given at Moriond, March 1996
\bibitem{eigen}
We thank G. Eigen for a discussion on this point.
\bibitem{distinguish}
I. Dunietz, FERMILAB-PUB-94/163-T [hep-ph/9409355]
\bibitem{yamamoto}
Hitoshi Yamamoto, private communication.
\bibitem{tagged}
I. Dunietz and J.L. Rosner, Phys. Rev. {\bf D34}, 1404 (1986);
I. Bigi et al., {\it in\/} CP violation, edited by C. Jarlskog
(World Scientific, Singapore, 1989), p.175
\bibitem{aleph}
D. Buskulic et al., ALEPH collaboration, CERN-PPE/96-30
\bibitem{BBL}
G. Buchalla, A.J. Buras and M.E. Lautenbacher,
"Weak Decays Beyond Leading Logarithms", FERMILAB-PUB-95/305-T,
{\it to appear in\/} Rev. Mod. Phys.
\bibitem{rhogamma}
A. Ali and C. Greub, Phys. Lett. {\bf B293}, 226 (1992);
Z. Phys. {\bf C60}, 433 (1993);
J.M. Soares, Phys. Rev. {\bf D49}, 283 (1994)
\bibitem{NIR2}
Y. Nir, Phys. Lett. {\bf B327}, 85 (1994)
\bibitem{BB}
G. Buchalla and A.J. Buras, Nucl. Phys. {\bf B412}, 106 (1994)
\bibitem{NES}
M. Neubert and C.T. Sachrajda, CERN-TH/96-19 [hep-ph/9603202]
\bibitem{LEE}
J. Lee-Franzini et al., Phys. Rev. Lett. {\bf 65}, 2947 (1990); 
P.~Abreu {\em et al.}, DELPHI collaboration, Zeit.~Phys. {\bf C68}, 
353 (1995)
 
\end{thebibliography}
\end{document}